# Electrically-Controlled Nuclear Spin Polarization and Relaxation by Quantum-Hall states

K. Hashimoto[1,2*], K. Muraki[1], T. Saku[1], and Y. Hirayama[1,2]

[1]NTT Basic Research Laboratories, NTT Corporation, 3-1 Morinosato-Wakamiya, Atsugi, Kanagawa 243-0198, Japan

[2]CREST-JST, 4-1-8 Honmachi, Kawaguchi, Saitama 331-0012, Japan

We investigate interactions between electrons and nuclear spins by using the resistance ($R_{xx}$) peak which develops near filling factor $\nu = 2/3$ as a probe. By temporarily tuning $\nu$ to a different value, $\nu_{temp}$, with a gate, the $R_{xx}$ peak is shown to relax quickly on both sides of $\nu_{temp} = 1$. This is due to enhanced nuclear spin relaxation by Skyrmions, and demonstrates the dominant role of nuclear spin in the transport anomaly near $\nu = 2/3$. We also observe an additional enhancement in the nuclear spin relaxation around $\nu = 1/2$ and $3/2$, which suggests a Fermi sea of partially-polarized composite fermions.

PACS numbers: 72. 90. +y, 75. 70. –i, 76. 60. –k.

Nuclear-spin systems and related phenomena in solids have attracted renewed attention in light of quantum computation because of their very long coherence time [1]. In metals, nuclear spins interact with conduction electrons mainly through contact hyperfine interactions [2]. The two subsystems can thereby exchange spin angular momentum via a simultaneous flip-flop, which forms the basis for dynamical nuclear polarization or spin-lattice relaxation at low temperatures. In quantum Hall (QH) systems, however, the quantization of the electrons' kinetic energy imposes a strong restriction on energy conservation. The Zeeman energy of electrons, which is about three orders of magnitude larger than that of the nuclei, must be compensated. As a result, flip-flop processes can take place only with the mediation of disorder [3]. Although current-driven nuclear spin polarization has been demonstrated for tunneling between spin-resolved edge channels [4], or near the onset of break down [5], these rely on continuous energy dispersion near the sample edge or a strong Hall electric field at large current.

The situation can be quite different in the extreme quantum limit. When electron correlation dominates over the single-particle Zeeman energy, the electronic spin system exhibits nontrivial behavior. The occurrence of a spin-unpolarized ground state for the Landau-level filling factor ($v$) of 2/3 is an example. Recently, an anomalous peak in the longitudinal resistance ($R_{xx}$) has been observed in the fractional QH (FQH) system near $v = 2/3$ in a GaAs/Al$_x$Ga$_{1-x}$As narrow quantum well (NQW) [6], and subsequently in a GaAs/Al$_{0.3}$Ga$_{0.7}$As single heterostructure [7]. Since the $R_{xx}$ peak developed over an unexpectedly long-time scale and showed a response in electrically-detected nuclear magnetic resonance (NMR), the authors of Ref. [6] suggested the involvement of nuclear spin polarization in the $R_{xx}$ anomaly. They argued that the nuclear-spin related $R_{xx}$ anomaly is associated with the presence of an electronic domain structure, which consists of the spin-polarized and spin-unpolarized $v = 2/3$-states and is formed around the transition point ($B_{2/3,t}$) between the spin states. On the other hand, a similar slow evolution of $R_{xx}$ observed for $v = 2/3$ [8] and 2/5 [9] has been discussed in a different context, in which the emphasis was placed on the domain dynamics of the electronic system itself, and it was claimed that the nuclear spins play only a secondary role.

In this Letter, we present magneto-transport data which demonstrates the electrical controllability of nuclear spin states in the QH regime. We study the relaxation of the $R_{xx}$ anomaly near $v = 2/3$ by temporarily tuning $v$ to other filling factors, $v_{temp} = 0$ (depletion) to 2, in order to turn off or modify the interactions between the electrons and nuclear spins with a gate. A striking enhancement of the $R_{xx}$ relaxation is observed on both sides of $v_{temp} = 1$, which arises from enhanced nuclear spin relaxation due to the Goldstone mode of the Skyrmions [10]. This demonstrates the crucial role of nuclear spins rather than the domain-wall dynamics of electronic systems for the $R_{xx}$ anomaly at $\nu = 2/3$. We also observe an additional enhancement in the nuclear relaxation around $\nu = 1/2$ and 3/2 which suggests a Fermi sea of partially-polarized composite fermions (CF's). Our experimental results open up a new way to realize all-electrical manipulation of nuclear spins in the solid state.

We use a back-gated GaAs/Al$_{0.3}$Ga$_{0.7}$As single heterostructure [11], which consists of a Si doped n-GaAs substrate, a 820-nm barrier layer, and a 500-nm GaAs top layer. We also study a gated GaAs/Al$_{0.3}$Ga$_{0.7}$As quantum well (QW) structure with a 20 nm thick well. The mobility of the single heterostructure (QW) is 220 (150) m$^2$/ Vs for an electron density ($N_s$) of 1.0 X 10$^{15}$ m$^{-2}$. We use a gate voltage to change the density of the two-dimensional electron gas (2DEG) and hence $\nu$. The samples are processed into 50 μm wide Hall bars. All the measurements are performed at a temperature, T < 150 mK except for the electrically-detected NMR experiment, which is done at 340 mK. $R_{xx}$ is measured using a standard low-frequency AC lock-in technique with a drain-source current, $I_{ds}$ = 10 – 100 nA.

In Fig. 1, $R_{xx}$ curves measured for different magnetic-field-sweep rates are compared for three different electron densities. The broken and solid curves are obtained by scanning the magnetic (*B*-) field downwards at a *normal* speed (0.2 T / minute) and a *slow* speed (0.003 T / minute), respectively. Figure 1(a) shows the results for the single heterostructure. For $N_s$ = 2.0 x 10$^{15}$ m$^{-2}$, the normal and slow speed scans give the same $R_{xx}$ curves. However, when we decrease $N_s$ to 0.89 x 10$^{15}$ m$^{-2}$, the two curves show different behavior. The difference is most pronounced on the low *B*-field side of $\nu = 2/3$, where $R_{xx}$ for the slow scan is enhanced by 30% compared to that for the normal scan. The enhancement increases to 50 % when $N_s$ is further decreased to 0.69 x 10$^{15}$ m$^{-2}$. We confirmed that it continues to grow as $N_s$ is reduced down to 0.46 x 10$^{15}$ m$^{-2}$. The results for the QW are shown in Fig. 1(b). A similar $R_{xx}$ enhancement is observed for $N_s$ = 1.2 x 10$^{15}$ m$^{-2}$. It disappears at higher density ($N_s$ = 1.6 x 10$^{15}$ m$^{-2}$) and at lower density ($N_s$ = 0.77 x 10$^{15}$ m$^{-2}$).

The anomalous $R_{xx}$ peak due to the $R_{xx}$ enhancement is clearly observed in the *B*-field region below 7 T and continuous to grow down to 3 T for the single heterostructure. On the other hand, for the 20-nm thick QW it is observed between 6 and 9 T and achieves maximum near 7 T. Note it was observed around 8 - 9 T for the 15-nm thick NQWs reported in Ref. [6]. These results can be understood by considering the different electron confinement in the single heterostructure and QWs, and furthermore a larger Zeeman (Coulomb interaction) energy favors the spin-polarized (unpolarized) $\nu = 2/3$ ground state. The thickness of the 2DEG in our single heterostructure sample is evaluated to be about twice that of the 15-nm thick NQW in Ref. [6]. This has the following two consequences. First, the Coulomb interaction energy is weakened. Second, the Zeeman energy is enhanced because of a larger $|g|$ value for the weaker confinement [12]. Both of these favor the spin-polarized ground state for $\nu = 2/3$ and so $B_{2/3,t}$ is shifted to lower B-field. The $R_{xx}$ enhancement therefore appears at lower $B$-field. Actually, electron spin-polarization measurements by photoluminescence indicate 2 T < $B_{2/3,t}$ < 3 T for a GaAs/Al$_x$Ga$_{1-x}$As single heterostructure [13]. This range is comparable to the $B$-field region in which we observe the largest enhancement for the single heterostructure. Note also that in Fig. 1(a) the most pronounced $R_{xx}$ enhancement is observed at a $B$-field slightly lower than that for exactly $\nu = 2/3$ ($B_{2/3}$). In contrast, it appears on the higher side of or at $B_{2/3}$ in Fig. 1(b) and in Ref. [6]. The position of $B_{2/3}$ with respect to $B_{2/3,t}$ determines whether the $R_{xx}$ enhancement appears to the higher or lower side of $B_{2/3}$.

In Fig. 2(a), we show the temporal evolution of $R_{xx}$ observed at a fixed $B$-field at $T$ = 140 mK for the single heterostructure. The measurement is initiated by first completely depleting the 2DEG for 7 hours and then setting $N_s$ = 0.92 X $10^{15}$ m$^{-2}$ ($\nu$ = 0.69) at $B$ = 5.8 T. For this $B$-field, the anomalous $R_{xx}$ peak clearly appears on the lower $B$-field side of the well-developed minimum in $R_{xx}$ at $\nu$ = 2/3. $R_{xx}$ develops with time ($t$) and saturates with an enhanced value ($\Delta R_{xx,s}$) of 1.6 k$\Omega$ after about 30 minutes. After the saturation, the sample is irradiated with a radio frequency wave and the frequency is swept upwards at a rate of 1.1 kHz / minute. As shown in Fig. 2(b), we observe a sharp drop in $R_{xx}$ when the frequency matches the resonance frequencies for the $^{75}$As, $^{69}$Ga and $^{71}$Ga nuclei, but not for $^{27}$Al. We have checked that $R_{xx}$ shows no response at other fractional filling factors, e.g. $\nu$ = 4/3, so we rule out any thermal effects. The absence of $^{27}$Al resonance is reasonable because the 2DEG distribution is mostly located (98 %) in the GaAs. Compared to the NMR signal of Ref [6] and [8], our signal shows faint fine structures. This may be due to the weaker 2DEG confinement in the heterostructure, although the mechanism is unclear.

The role of nuclear spins is further highlighted in the following relaxation measurement. After $R_{xx}$ saturates at $\nu$ = 0.69, we completely deplete the 2DEG for a given period of time ($\tau$) of 30 s. We then restore the same 2DEG density and measure $R_{xx}$ again [Fig. 2(a)]. Surprisingly, $R_{xx}$ does not relax to the initial value obtained at $t$ = 0 s. Rather, it returns to the value just before depletion. Here, the resistance relaxation ($\Delta R_{xx}$) is only 18 % of the initial enhancement (1.6 k$\Omega$). We measure $\Delta R_{xx}$ for several values of $\tau$ [see inset of Fig. 2(a)]. By fitting the data to the form $\Delta R_{xx}(\tau) = \Delta R_{xx,s}$ [1 -

exp($-\tau/T_r$)], we find that the rate of the resistance relaxation is as small as $1/T_r = 10^{-3}$ s$^{-1}$.

The above results clearly demonstrate that the cause of the anomalous $R_{xx}$ peak is memorized by the nuclear spin system, but not in the 2DEG itself. We speculate that the $R_{xx}$ peak is reproduced in the following mechanism. First, nuclear spins are locally polarized with the flip-flop process driven by the current flow across electronic domain walls. The nuclear spin polarization locally changes the effective Zeeman splitting of the electrons by the contact hyperfine interaction. It affects the domain morphology and hence $R_{xx}$. Since the inhomogeneous hyperfine field defines different potential landscapes for the spin-polarized and the spin-unpolarized $\nu = 2/3$-states, the domain structure is reproduced when the 2DEG is induced again. The small but finite relaxation during the 2DEG depletion may be mainly governed by nuclear spin diffusion. The value of $1/T_r = 10^{-3}$ s$^{-1}$ obtained here is comparable to the nuclear spin diffusion rate in high-purity bulk GaAs measured directly in nuclear spin relaxation measurements [14]. This justifies our simple assumption that the $R_{xx}$ enhancement is proportional to the degree of nuclear spin polarization, i.e. $1/T_r = 1/T_1$.

We now address how nuclear spin relaxation is modified in the presence of a 2DEG at various filling factors. The measurements are carried out at a constant $B$-field of 5.8 T. After waiting for $R_{xx}$ to saturate at $\nu = 0.69$, the filling factor is quickly changed to a temporary value ($\nu_{temp}$) using the back gate, and held at $\nu = \nu_{temp}$ for a given period of time $\tau$. Subsequently, the filling factor is restored to $\nu = 0.69$ to measure the resistance relaxation ($\Delta R_{xx}$) during the time $\tau$. Representative results for three values of $\nu_{temp}$ (= 0.47, 0.95 and 2.21) are plotted as a function of $\tau$ in Fig. 3(a). The case for complete depletion ($\nu_{temp} = 0$) is also shown for reference. In Fig. 3 (b), $1/T_1$ obtained by curve fitting is plotted for various values of $\nu_{temp}$ between 0.33 and 2.3. On either side of $\nu_{temp} = 1$, $1/T_1$ is enhanced strongly and reaches 0.4 s$^{-1}$ which is two orders of magnitude faster than the value for the case of depletion.

This significantly enhanced relaxation can be explained by considering the existence of Skyrmions [15], which appear either side of $\nu = 1$. The nuclear spins are efficiently flipped by the low-frequency spin fluctuation associated with the Goldstone mode of Skyrmions [10]. The fast nuclear spin-relaxation on either side of $\nu = 1$ has been observed by optically pumped NMR (OPNMR) [16], and by specific heat measurements [17]. It is noteworthy that our result, $1/T_1 = 0.4$ s$^{-1}$, is one order of magnitude faster than the OPNMR result. This may be due to a larger Skyrmion size which would efficiently relax the nuclear spin polarization. The size of the Skyrmion can be larger, because we use lower $B$-fields than those in Ref. [16]. In addition, it is also possible that "clumped" Skyrmions [18] localize around the polarized nuclear spins, which can then efficiently relax the nuclear spin polarization. Such Skyrmion localization may arise from the inhomogeneous hyperfine field caused by the nuclear spin polarization [19], or by an inhomogeneous electrostatic potential due to large length scale density fluctuations [18].

Finally, a closer examination of the data reveals that additional enhancements of $1/T_1$ exist around $v_{temp}$ = 1/2 and 3/2, where $1/T_1$ reaches $10^{-2}$ s$^{-1}$ [see dashed lines in Fig. 3(b)]. One might expect the rate for disorder-mediated flip-flop processes to be governed by the magnitude of $\sigma_{xx}$ [3]. While the $v_{temp}$ dependence of $1/T_1$ resembles that of $\sigma_{xx}$ [inset of Fig. 3(b)], the difference between is apparent. There is no peak in $1/T_1$ between $v$ = 5/3 and 2, whereas a clear peak exists in $\sigma_{xx}$. An interesting possibility is that the enhanced relaxation is due to the presence of a Fermi sea of partially-polarized CF's at $v$ = 1/2 and 3/2 [13]. The continuous density of states at the Fermi surface of partially-polarized CF's allows a flip-flop process in which the excess Zeeman energy of a CF is transferred to the kinetic energy of the CF. At *B*-field far away from $v$ = 1/2 (or 3/2), the density of states of CF's is quantized by the effective *B*-field, $B^* = B - B_{1/2}$ ($B_{1/2}$: *B*-field for $v$ = 1/2) [20], which then hampers the flip-flop process. In this picture, the enhanced electron-nuclear spin interactions near $v$ = 2/3 can be interpreted as due to the coincidence of CF Landau levels, where the excess Zeeman energy of a CF is compensated by the cyclotron energy of the CF.

In conclusion, we have shown that nuclear spins are polarized in the QH system near $v$ = 2/3, and the resulting $R_{xx}$ enhancement can be used to measure the nuclear spin polarization. By performing relaxation measurements, we observe extremely fast relaxation on either side of $v$ = 1. The fast relaxation can be explained in terms of the Goldstone mode of Skyrmions. The enhancements of $1/T_1$ around $v_{temp}$ = 1/2 and 3/2 may be due to the Fermi sea of partially-polarized CF's. Our results demonstrate that by appropriately tuning the filling factor in a QH system, one can electrically create, detect, and destroy nuclear spin polarization in the solid state.

The authors acknowledge T. Fujisawa S. Miyashita, G. Austing, Y. Tokura and A. Khaetskii for fruitful discussion and support.

References

[1] B.E. Kane, Nature **393**, 133 (1998).
[2] A. Abragam, The principles of Nuclear Magnetism (Oxford, 1961).
[3] A. Berg *et al*., Phys. Rev. Lett. **64,** 2563 (1990).
[4] K.R. Wald, L.P. Kouwenhoven, and P.L.McEuen, Phys. Rev. Lett. **73**, 1011 (1994); B.E. Kane, L.N. Pfeiffer, and K.W. West, Phys. Rev. B **46**, 7264 (1992).
[5] A.M. Song and P. Omling, Phys. Rev. Lett. **84**, 3145 (2000).
[6] S. Kronmüller *et al*., Phys. Rev. Lett. **81**, 2526 (1998); *ibid*. **82**, 4070 (1999).
[7] K. Hashimoto *et al*., Physica B **298**, 191 (2001); K. Hashimoto, T. Saku, and Y. Hirayama, Proceedings of the International Conference on the Physics of Semiconductors, M148 (2000).
[8] J.H. Smet *et al*., Phys. Rev. Lett. **86**, 2412 (2001).
[9] J. Eom *et al*., Science **289**, 2320 (2000).
[10] R. Côté *et al*., Phys. Rev. Lett. **78**, 4825 (1997).
[11] U. Meirav, M. Heiblum, and F. Stern, Appl. Phys. Lett. **52**, 1268 (1988); Y. Hirayama, K. Muraki, and T. Saku, Appl. Phys. Lett. **72**, 1745 (1998).
[12] R.M. Hannak *et al*., Solid State Communication. **93**, 313 (1995).
[13] I.V. Kukushkin, K.v. Klitzing, and K. Eberl, Phys. Rev. Lett. **82**, 3665 (1999).
[14] J.A. McNeil and W. G. Clark, Phys. Rev. B **13**, 4705 (1976).
[15] S.E. Barrett *et al*., Phys. Rev. Lett. **74,** 5112 (1995).
[16] R. Tycko *et al*., Science **268**, 1460 (1995).
[17] V. Bayot *et al*., Phys. Rev. Lett. **76**, 4584 (1996).
[18] P. Khandelwal *et al*., Phys. Rev. Lett. **86,** 5353 (2001).
[19] I.V. Kukushkin, K.v. Klitzing, and K. Eberl, Phys. Rev. B **60**, 2554 (1999).
[20] R. R. Du *et al*., Phys. Rev. Lett. **70,** 2944 (1994).

Figure captions

Fig. 1 *B*-field dependence of the $R_{xx}$ enhancement near $\nu = 2/3$ with different electron densities tuned by the back-gate bias. The broken and solid curves, respectively, show $R_{xx}$ measured with a normal scan speed (0.2 T / minute) and a slow scan speed (0.003 T / minute). All data is acquired when the *B*-field is scanned downwards. (a) A gated single heterostructure for $T = 100$ mK and $I_{ds} = 100$ nA. (b) A gated QW for $T = 80$ mK and $I_{ds} = 10$ nA.

Fig. 2 Development and relaxation of anomalous $R_{xx}$ peak at $B = 5.8$ T for $N_s = 0.92 \times 10^{15}$ m$^{-2}$ ($\nu = 0.69$). (a) Time dependence of $R_{xx}$ at T = 140 mK. $R_{xx}$ saturates after about 30 minutes. Next, the 2DEG is completely depleted for 30 s and subsequently is refilled again. $\Delta R_{xx}$ represents the relaxation value of the $R_{xx}$ enhancement. Inset shows $\Delta R_{xx}$ versus the 2DEG depletion time, $\tau$. The dashed line is the fitted curve. (b) Electrically-detected NMR of the $^{75}$As, $^{69}$Ga and $^{71}$Ga nuclear spins at T = 340 mK. The sample is irradiated with a radio frequency wave after the $R_{xx}$ enhancement saturates.

Fig. 3 Relaxation measurement in the presence of the 2DEG at B = 5.8 T and T = 140 mK. (a) $\tau$ dependences of $\Delta R_{xx}$ for three typical temporary filling factors ($\nu_{temp}$). Here, $\Delta R_{xx}$ represents the relaxed $R_{xx}$ value at $\nu = 0.69$ ($N_s = 0.92 \times 10^{15}$ m$^{-2}$) when the filling factor is set to $\nu_{temp}$ for a given period of time, $\tau$. The case for complete depletion of the 2DEG is also shown for reference. The dotted lines are fitted curves which give the relaxation rate $1/T_1$. (b) $1/T_1$ versus $\nu_{temp}$. For $\nu_{temp} < 0.7$ and $\nu_{temp} > 1.3$, the dashed lines shows the curve enlarged by a factor of 10. Inset: $\nu$ dependence of the conductivity $\sigma_{xx}$, which is calculated from the measured resistance tensor.

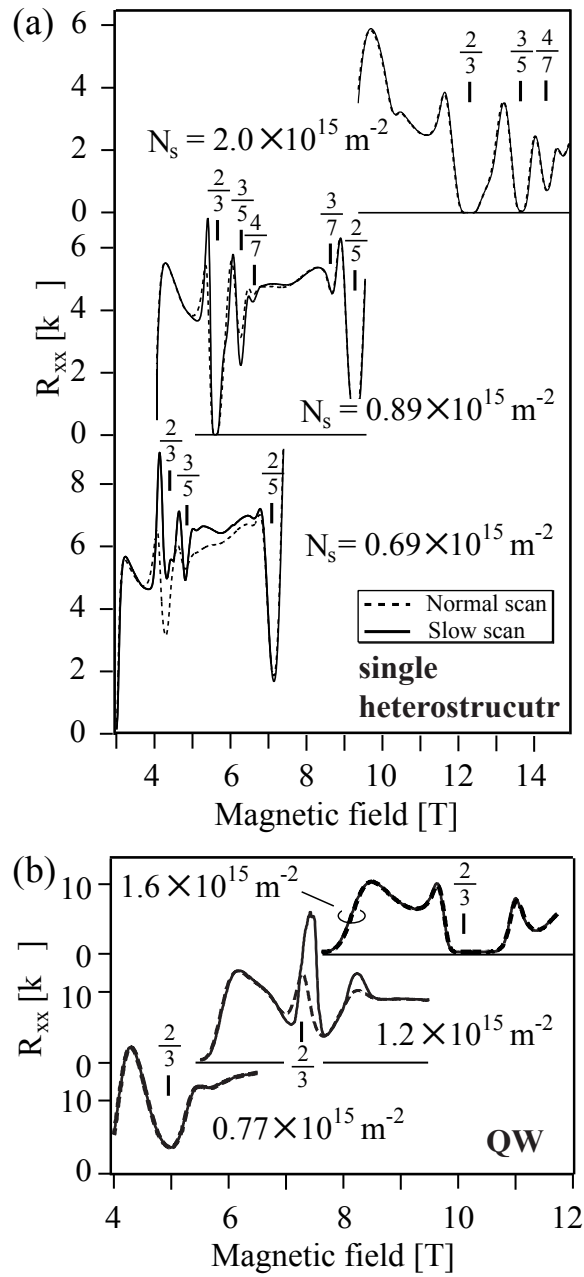

Fig. 1 Hashimoto et al.

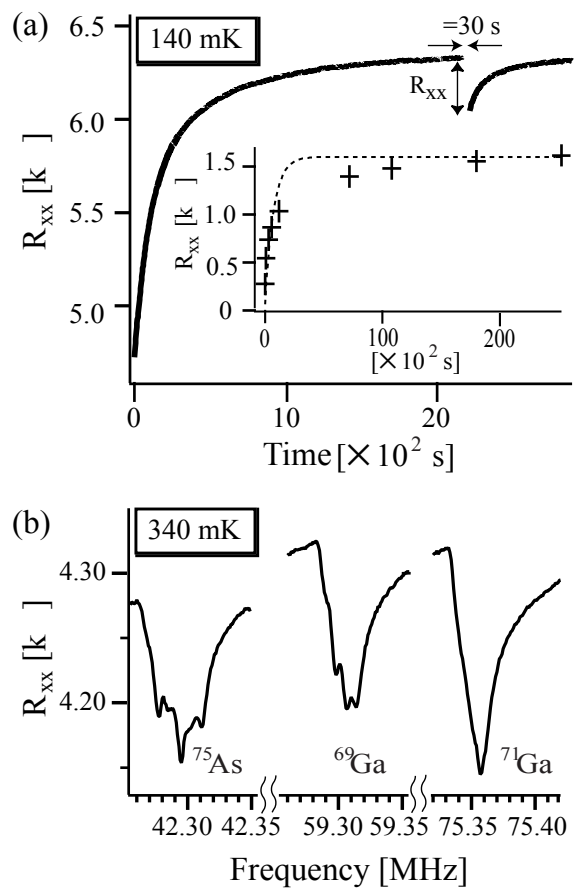

Fig. 2 Hashimoto et al.

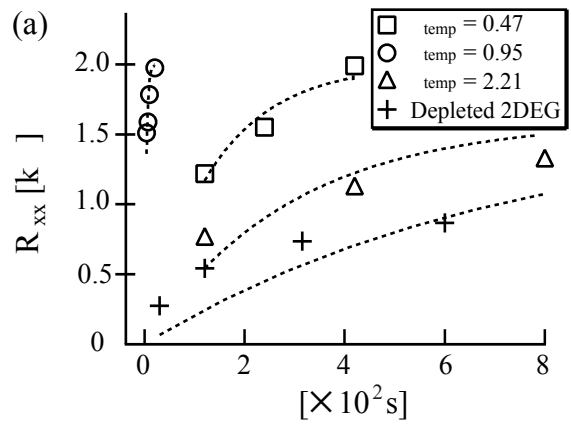
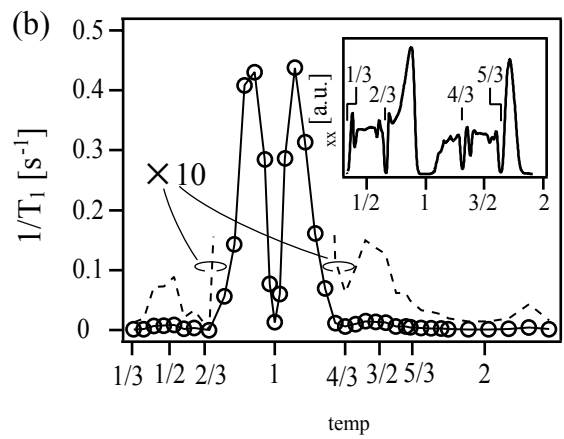

Fig. 3 Hashimoto et al.